\documentstyle[twoside,fleqn,espcrc2,epsfig]{article}

\newcommand{\be}{\begin{equation}}
\newcommand{\ee}{\end{equation}}

\newcommand{\AmS}{{\protect\the\textfont2
  A\kern-.1667em\lower.5ex\hbox{M}\kern-.125emS}}

\centerline {\bf ~~~~~~~~~~NOE: a neutrino experiment for the 
CERN-Gran Sasso Long Base line project}
\centerline {\bf ~~~~~~~~~~~~~~~~~E. Scapparone for the NOE Collaboration}
\centerline{~~~~~~~~~~ Laboratori Nazionali del Gran Sasso}
\centerline{~~~~~~~~~~ S.S.17 km 18+910, 61070, Assergi (AQ), Italy}      
\centerline{ \bf ~~~~~~~~~~Talk at 5th International Workshop on Tau Lepton Physics (TAU98)}
\centerline{ \bf ~~~~~~~~~~Santander, Spain, September 14 - 17, 1998} 
\centerline{\bf ~~~~~~~~~~To be published in Nucl. Phys. B, Proc. Suppl} 
\centerline{\bf ~~~~~~~~~~$($ writeups available at 
http$://$www.ifca.unican.es/tau98/index$\_$writeups.html$)$}          

\title{NOE: a neutrino experiment for the CERN-Gran Sasso Long Base line project}

\author{E. Scapparone\address{INFN, LNGS
        S.S.17 km 18+910, 61070, Assergi (AQ), Italy}
       for the NOE Collaboration\thanks{see [5] for the complete list
of the Collaboration}
       }

\begin{document}

\begin{abstract}
The project of a large underground experiment (NOE) devoted to long 
baseline neutrino oscillation measurement is presented. The apparatus,
consisting of TRD and calorimeter modules, 
has been optimized to be sensitive in the region of $\sin^2 2\theta$ 
and $\Delta m^2$ suggested by the atmospheric neutrino oscillation 
signal.

\end{abstract}\vspace{0.2cm}

\maketitle
\section{Introduction}

The scientific goal of the NOE long baseline (LBL) experiment is the 
measurement of neutrino masses looking at $\nu_\mu \rightarrow \nu_e$ 
and $\nu_\mu \rightarrow \nu_\tau$ oscillations. The philosophy of NOE 
design is to have oscillation sensitivity by looking for the $\tau$ 
decay ($\nu_\mu \rightarrow \nu_\tau$ oscillation) or for an electron 
excess ($\nu_\mu \rightarrow \nu_e$ oscillation) and by measuring a 
deficit of muons in apparent NC/CC ratio.

The main experimental hint for the $\nu$ oscillation search in the 
region of low $\Delta m^2$ ($10^{-2} \div 10^{-3} \ eV^2$) comes from 
the muon deficit observed in atmospheric neutrino flux measurements 
\cite{soudan,macro1,superk}. Recent results from LBL reactor 
neutrino experiment (CHOOZ) excluded neutrino oscillation in $\bar\nu_e$ 
disappearance mode, up to $\sin^2 2\theta > 0.18$ for large $\Delta m^2$
\cite{chooz}.

Taking into account the confirmations of the atmospheric neutrino anomaly and the 
negative CHOOZ result, a LBL experiment has to fulfill the 
following requirements:

\begin{enumerate}
\item {\bf $\nu_\tau$ tagging.} 
The search for $\nu_\tau$ appearance is mandatory
to confirm the oscillation phenomenon. This search requires 
detector high performances to tag the $\tau$ decay.

\item {\bf Measurement of the ratio NC/CC.} This robust and 
unambiguos test is important to investigate on the existence of a 
neutrino oscillation signal. 
There is no doubt that 
this measurement can be done only with a massive detector.

\item {\bf  Atmospheric neutrinos.} After the last results from
Superkamiokande, suggesting smaller values of $\Delta m^2$, the interest for the 
atmospheric neutrinos is raised up. It would be interesting to test
this effect using a massive apparatus based on a different technique 
with respect to the water \v{C}erenkov detectors.

\item {\bf Fast response.} If a beam from CERN to Gran Sasso will be
available in the next years, a strong competition with American and 
Japanese LBL programs is foreseen: at present, the 7 kton NOE project 
can adequately compete with the 8 kton MINOS detector and with K2K.
\end{enumerate}

According to these remarks the NOE program can be summarized 
in this way:

\begin{itemize}
\item Direct $\nu_\tau$ appearance by kinematical $\tau$ decay 
reconstruction and inclusive (NC/CC) $\nu_\mu$ disappearance.

\item Investigation of $\nu_\mu \rightarrow \nu_e$ oscillation in
a mixing angle region two orders of magnitude beyond the CHOOZ limit.

\item Atmospheric neutrino studies.
\end{itemize}

In order to improve the $\nu_\tau$ search, the apparatus has been 
equipped with Transition Radiation Detector (TRD) interleaved 
between calorimetric modules (CAL). The combination of TRD and CAL 
information strongly enforces $e$, $\mu$, and $\pi$ identification, 
thus permitting the study of the $\tau$ decay. In particular the 
$\tau$ appearence can be detected
through the $\tau \rightarrow e\nu\nu$ channel,
with a clean 
signature, taking advantage of the low background (residual $\nu_e$ beam). 
Moreover the good electron identification in the TRD and the low 
$\pi^o$ background allow to reach high $\sin^2 2\theta$ sensitivity 
looking for $\nu_\mu \rightarrow \nu_e$ oscillation, thus considerably 
enlarging the region investigated by CHOOZ.

\section{The NOE detector}
\begin{figure}
\hspace{2cm}
\vskip 1cm
\epsfig{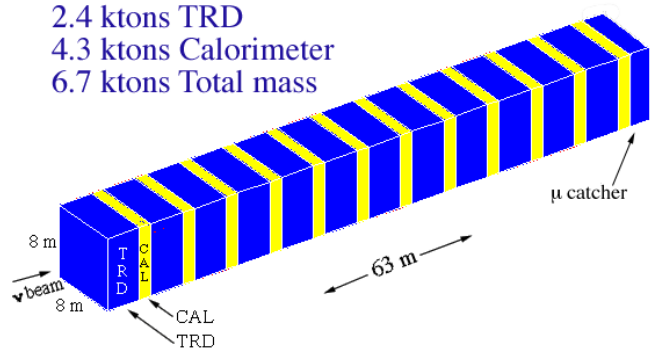}
\vspace{1cm}
\caption{\it Schematic view of the NOE detector \label{fig:noex}} 
\end{figure}

\begin{figure}
\hspace{3cm}
\vskip 2cm
\epsfig{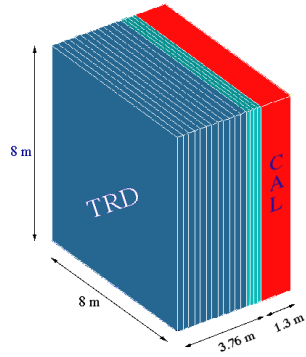}
\vspace{-1cm}
\caption{\it The Basic Module of the NOE detector \label{fig:detm}} 
\end{figure}

The detector (Fig.~\ref{fig:noex}) is a typical fixed target 
apparatus consisting of a sequence of 12 basic modules(BM). 
Each BM is composed 
by a lighter part (TRD target) in which vertex, $e$-identification and
kinematics are defined, followed by a scintillating fiber calorimeter 
devoted to energy measurement and event containement. The Basic 
Module (BM) is shown in Fig.~\ref{fig:detm}. Appearance measurement
is performed using events generated in TRD ($2.4 \ kton$), while
disappearance measurements are performed by looking at events generated
both in the TRD and in the calorimeter (total mass $\sim 7 \ kton$).


The calorimetric element is an 8 meter long bar filled with
iron ore and scintillating fibers\footnote{Extruded scintillator 
strips with wavelength shifter fibers have been also studied.}. 
The calorimetric 
module is made by alternate planes of crossed bars. The calorimetric 
bar consists of more logical cells with square cross section. 
The iron ore is radiopure and practically cost free.

A lot of R$\&$D has been performed to improve the scintillating fibers
performance.
The present production of $2 \ mm$ diameter
scintillating fibers provide an attenuation length $\lambda$=$4.5 \ m$
and an light yield L $\geq$ 20 pe/MeV. These 
figures allow to build $8 \ m$ long bars. Further investigations 
to improve fiber features are in progress: longer fibers
would allow to increase the NOE cross section ($9 \times 9 \ m^2$) 
and therefore the total mass ($8 \ kton$). It is worth noting the 
very high intrinsic granularity of the proposed calorimeter : the 
average distance between the fibers inside the absorber is of the 
order of $3 \ mm$. The fibers are 
grouped together at each side of the calorimetric bar and sent to 
single or multipixel photodetector.

The energy resolutions for electrons and hadrons in the calorimeter
have been evaluated by means of a GEANT Montecarlo simulation. They are, 
respectively, $\sigma (E)/E=0.01 + 0.17/ \sqrt{E}$ and 
$\sigma (E)/E=0.08 + 0.42/ \sqrt{E}$.
The TRD module consists of 32 vertical layers of $8 \times 8 \ m^2$ 
area, each made by polyethylene foam radiator ($\rho \sim 100 \ mg/cm^3$) 
and 256 proportional tubes ($3 \times 3 \ cm^2$ cross section), 
filled with an $Ar$ (60\%) - $Xe$ (30\%) - $CO_2$ (10\%) mixture, 
already tested in MACRO experiment. Consecutive layers have tubes 
rotated of $90^\circ$.

A graphite wall of $5 \ cm$ thickness is set in front of each of the
first 24 layers of the TRD module acting as a $174 \ ton$ target for 
$\nu_e$ and $\nu_\tau$ interactions, to be identified in the following 
layers. The last target wall is followed by 8 TRD layers in order 
to identify the secondary particles. Each target wall corresponds to 
$0.25 \ X_0$ while the entire TRD basic module corresponds to about 
$7 \ X_0$ and $3.5 \ \lambda_I$. The total length is about $3.76 \ m$.

So many layers of proportional tubes permit to determine the muon 
energy by means of multiple measurements of energy loss $dE/dx$.
Combining the informations coming from both subdetectors (TRD and CAL) the 
discrimination between $e$, $\mu$ and $\pi$ is largely enforced allowing 
the study of several neutrino oscillation channels.

\section{$\nu_\tau$ appearance and requirements about $\nu$ beam}
The rate of $\nu_\tau$ CC events is given by
\be R_\tau = A \int \sigma_\tau P_{osc} \Phi dE, \ee
where E is the energy, $\sigma_\tau$ the $\nu_\tau$ CC cross section, 
$P_{osc}$ the oscillation probability, $\Phi$ the muon neutrino flux
and $A$ the number of target nucleons in the detector. The search for 
$\nu_\tau$ requires that the term $\sigma_\tau P_{osc} \Phi$ is large. 
Therefore a dedicated $\nu$-beam has to provide most of its flux in 
the energy range where the factor $\sigma_\tau P_{osc}$ is larger.

\begin{figure}
\epsfig{file=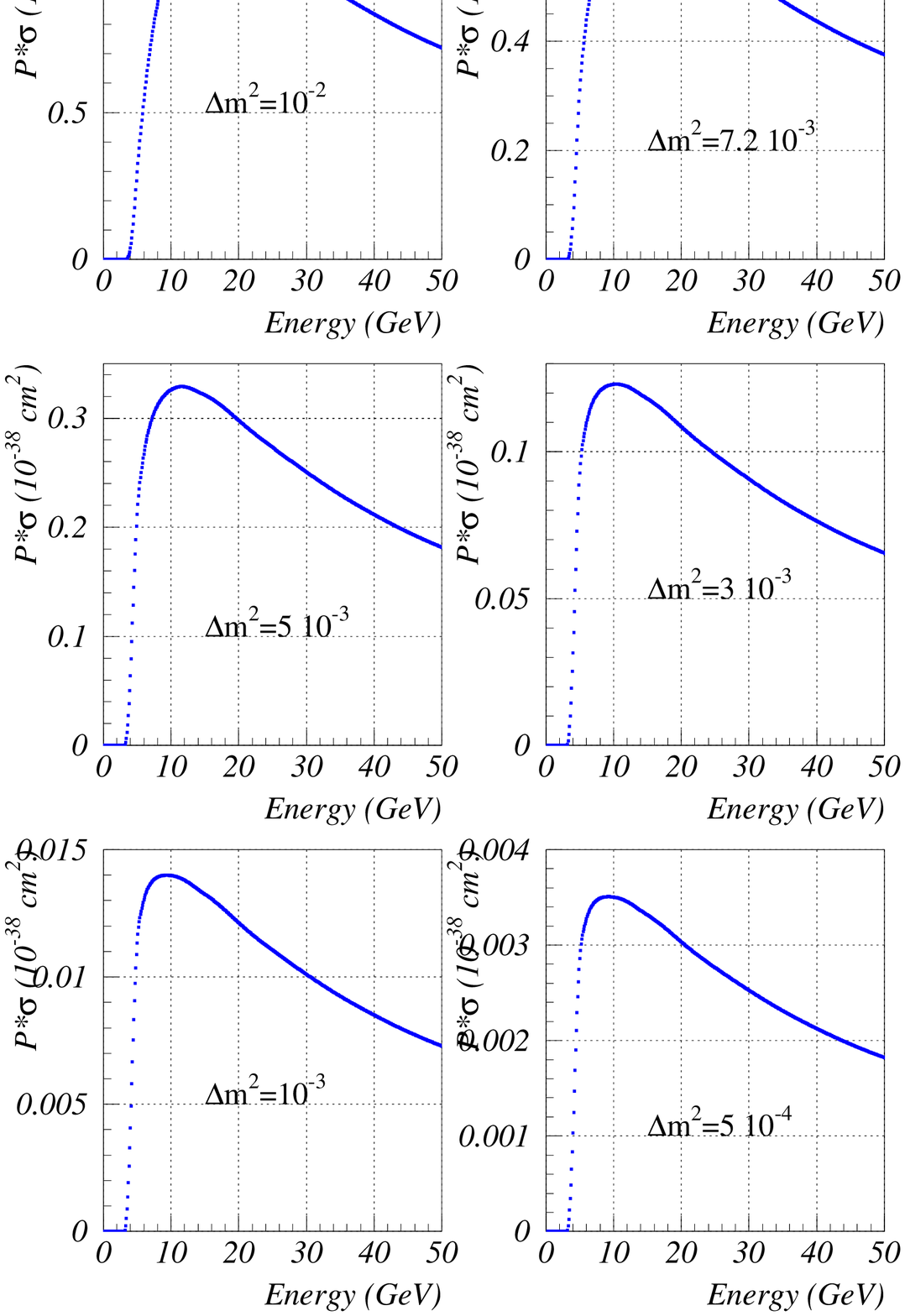,width=11cm,height=11cm,bbllx=10bp,bblly=0bp,bburx=770bp,bbury=770bp}
\vskip -3.cm
\caption{\it \label{fig:p_per_sigtau} $\nu_\tau$ CC cross section 
        multiplied by the oscillation probability for 
        different values of $\Delta m^2$.}
\end{figure}

Assuming the mixing of two neutrinos, the oscillation probability is
\be P_{osc} = \sin^2 2\theta \sin^2 (1.27 \Delta m^2 L/E), \ee
where $L = 731 \ km$ is the distance CERN - Gran Sasso.
We have to take into account that the $\tau$ cross 
section grows slowly with energy above a threshold of about $3.5 \ GeV$. 
The factor $\sigma_\tau P_{osc}$ is shown in 
Fig.~\ref{fig:p_per_sigtau} for different values of $\Delta m^2$. The 
optimal energy is about $15 \ GeV$ for $\Delta m^2 = 0.01 \ eV^2$ and 
decreases gently with $\Delta m^2$ towards a limiting value of about 
$10 \ GeV$.


In the following, the beam from Ref.~\cite{ball} and 5 years of data
taking are assumed.
\begin{figure}
\vskip 0.7cm
\psfig{file=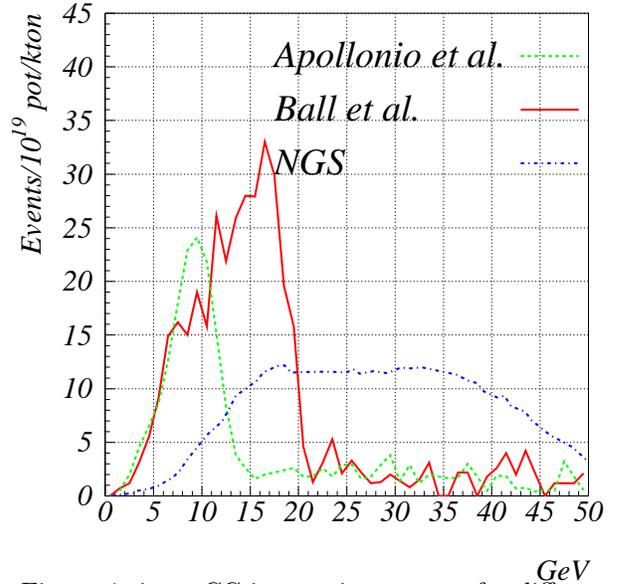,width=8cm,bbllx=31bp,bblly=180bp,bburx=550bp,bbury=630bp}
\vskip -3.7cm
\caption{it \label{fig:spettri} $\nu_\mu$ CC interaction spectra
        for different proposed beams.}
\vskip -1.2cm
\end{figure}
\section{$\tau$ appearance searches}
Tau appearance search is performed on the basis of kinematical
identification of the $\tau$ decay. The $\tau \rightarrow e \nu \nu$ 
is the favourite channel for this search due to the low background 
level and the the good electron identification capability of the TRD.
It is worth noting that in the region of atmospheric anomaly the
oscillation probability is $50 \div 100$ times higher than expected in
NOMAD, as a consequence a much lower background rejection power is
required.

In order to check the overall NOE performances, a complete chain of
event simulation and analysis has been performed.
Event generators that include Fermi motion, $\tau$ polarization and
nuclear rescattering inside the nucleus have been used to simulate
quasi elastic, resonance and deep inelastic events.

Generated events are processed by a GEANT based MonteCarlo in which 
calorimeter and TRD geometrical set-up are described in detail,
down to a scale of a few $mm$. Fiber attenuation length, Birks 
saturation, photoelectron fluctuations and readout electronics non 
linearities for both TRD and calorimeter have been taken into account.
DST of processed events ($\tau \rightarrow e \nu \nu$, $\nu_\mu$ NC
and $\nu_e$ CC) have been produced and analysed.

Electron identification is performed by looking for high energy releases in
the TRD and in the calorimeter readout elements in fully contained events. 
The electron candidate is the one that maximizes collected energy in 
a $5^\circ$ cone centered at the interaction vertex. Electron direction 
is reconstructed by weighting hits position by collected energy.
With present algorithms an angular resolution of $0.6^\circ$ and a 
$180 \ MeV/c$ resolution on the measurement of transverse momentum are 
achieved.

The remaining part of the event is used to reconstruct the hadronic
component. The obtained resolution on the measurement of transverse
momentum is $420 \ MeV/c$.
 
\begin{figure}
\epsfig{file=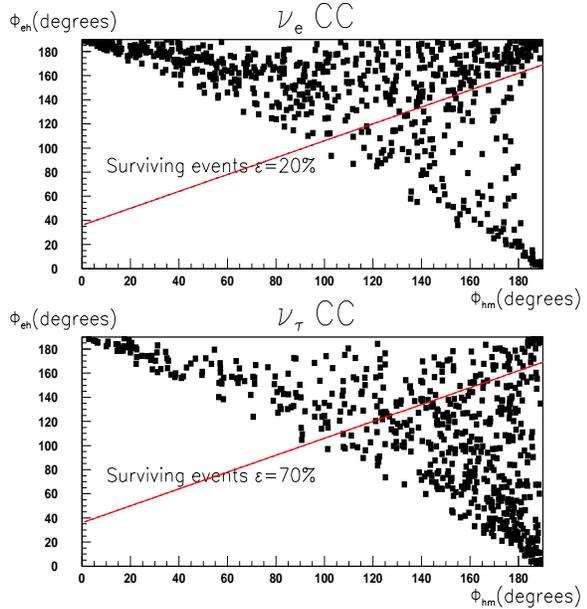,width=8.5cm,bbllx=31bp,bblly=150bp,bburx=550bp,bbury=630bp}
\vspace{-0.5cm}
\caption{\it \label{fig:fifi} Effects of $\phi_{e-h}$ $\phi_{m-h}$ cut 
on background ($\nu_e$) and signal ($\nu_\tau$).}
\end{figure}

\begin{figure} [t]
\epsfig{file=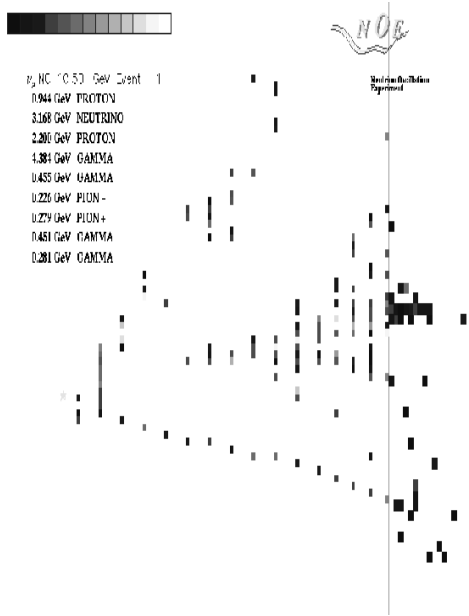,height=10.cm,width=7.5cm}
\vskip -1.5cm
\caption{\it \label{fig:bandiera} The event display of a typical NC event with 
a $\pi^{o}$. The hits on the left of the vertical line belong to the TRD, those 
on the right belong to the Calorimeter.}
\vskip -0.5cm
\end{figure}

Topological cuts on the electromagnetic shower are applied to reject 
$\nu_\mu$ NC events with $\pi^o$ faking electrons. Fig.\ref{fig:bandiera}
shows a tipical
neutral current event with a $\pi^{o}$: the electromagnetic shower 
doesn't start in the event vertex, allowing an easy rejection of the 
event. Work is in progress 
to improve the reconstruction efficiency and $\nu_\mu$ NC rejection.
Additional cuts are performed to reduce the background:

\begin{itemize}
\item total reconstructed energy $< 15 \ GeV$,
\item electron energy $> 1.5 \ GeV$,
\item the component of electron momentum perpendicular to the hadronic
      jet direction $Q_{lep} > 0.75 \ GeV/c$,
\item transverse mass\vskip 1mm 
      $M_T = \sqrt{4p_T^ep_T^m \sin^2(\phi_{e-m}/2)} < 2 \ GeV$,
\item $\phi_{e-h}$ $\phi_{m-h}$ correlation as shown in 
      Fig.~\ref{fig:fifi}.
\end{itemize}
Resulting efficiencies and residual backgrounds are reported in 
Table~\ref{tab:effi}.
\begin{table}[t]
\epsfig{file=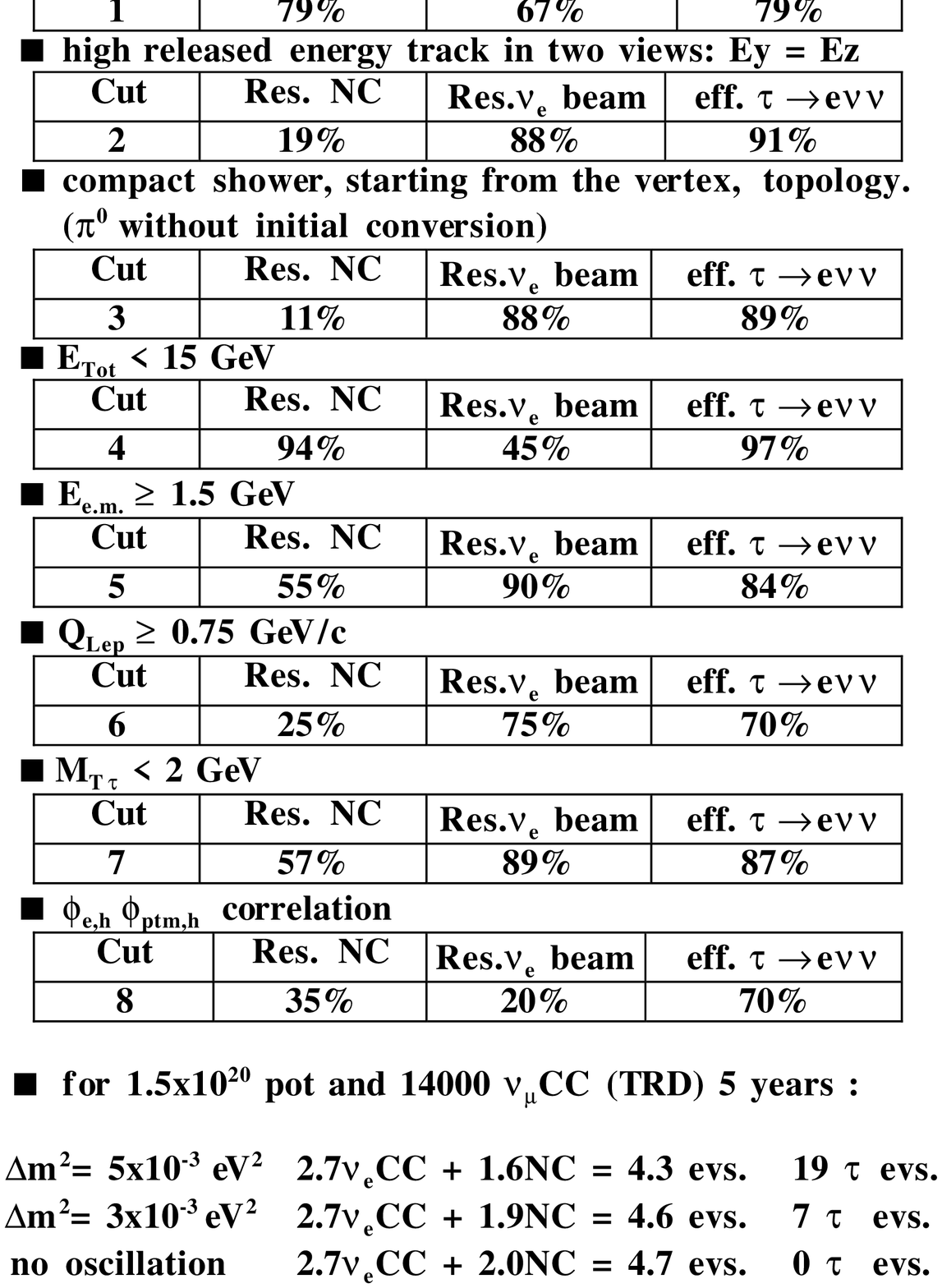,width=7.5cm,height=12cm}
\vskip -0.5cm
\caption{\it \label{tab:effi}Details of analysis cuts, efficiency and rejection
power for main background sources.}
\vskip -0.5cm
\end{table}     
\section{Neural network for the ratio NC/CC}
\begin{figure} [t]
\vskip -4cm
\epsfig{figure=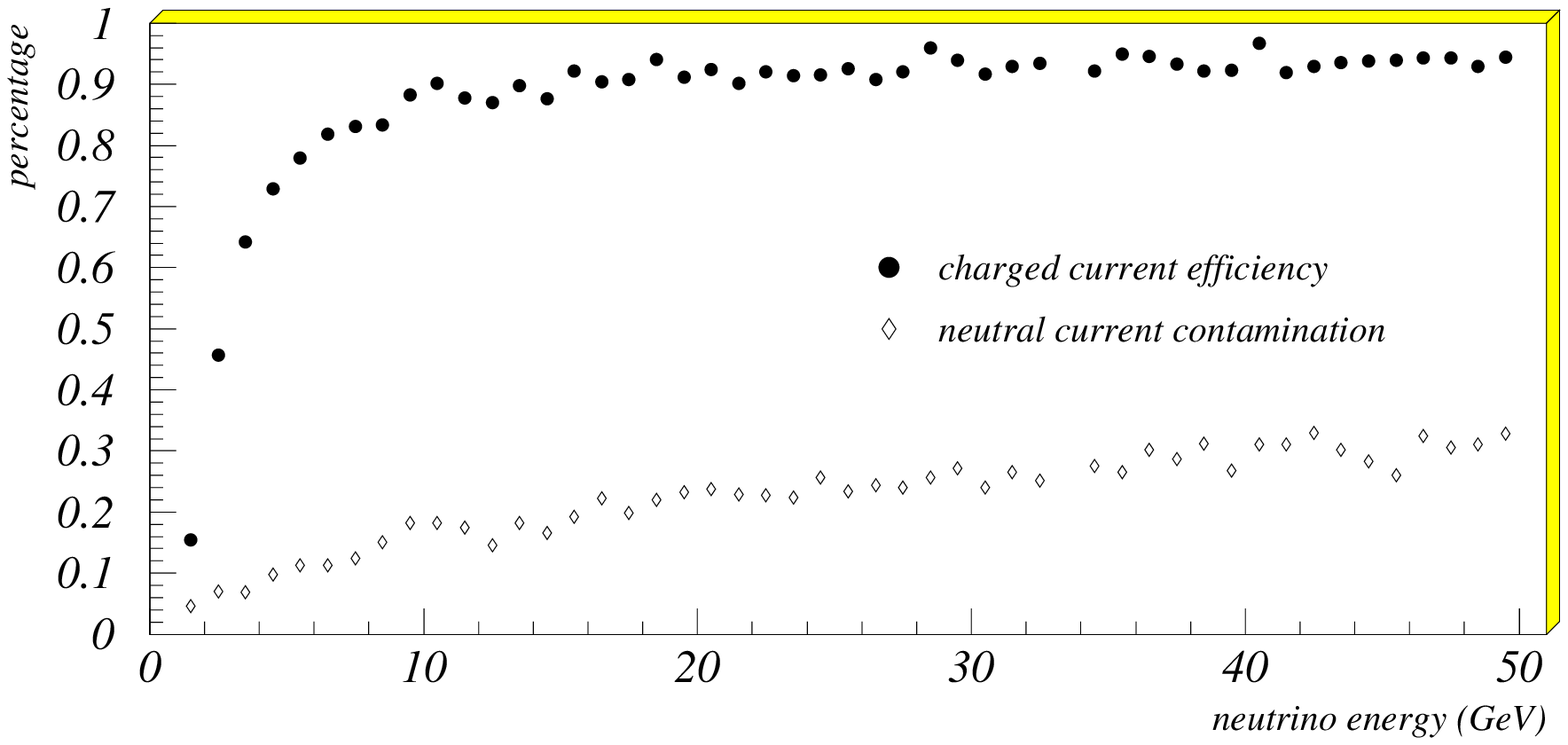,width=8.5cm}
\vskip -8cm
\epsfig{figure=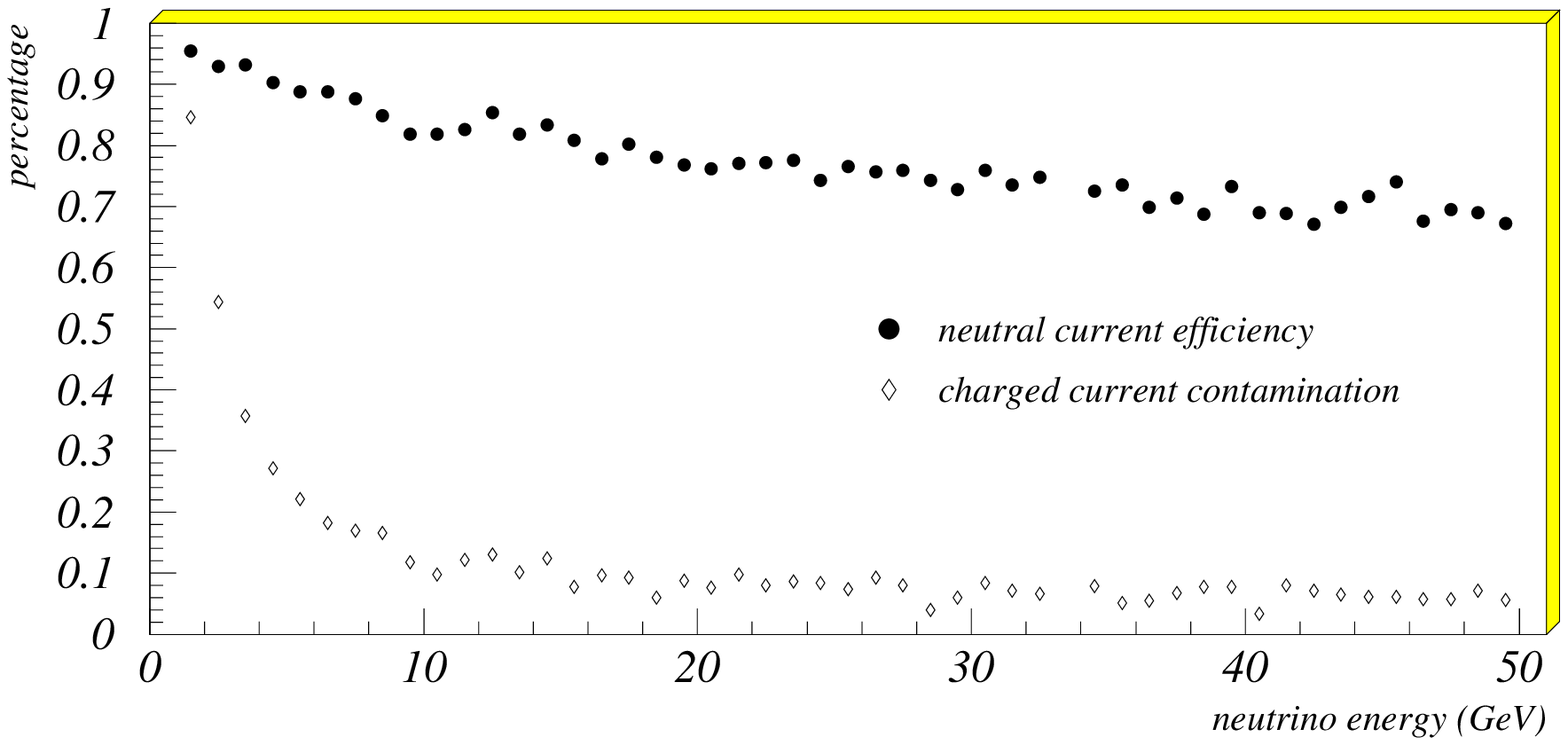,width=8.5cm}
\vskip -5cm
\caption{\it CC and NC signal. The contamination is also shown. 
  \label{fig:orlo}} 
\end{figure}

The measurement of the NC/CC ratio is performed looking at the 
observable $R_{obs}$= (no $\mu$)/$\mu$, where (no $\mu$) is the number
of events not showing a muon track while $\mu$ is the number of events with
a reconstructed muon. 
The procedure to perform this measurement is well known \cite{noe6}. 
The oscillation probability can be written as:
\be
P=\frac{R_{obs}-R_{th}-\epsilon(R_{th}+1)}{R_{obs}(1-{\eta}B)+\eta(1-B)}
\ee
where $\epsilon$ is the ratio of $\nu_{e}$ to $\nu_{\mu}$ in the beam, 
B is the branching ratio
for $\tau$ decay in muon, $R_{th}$=$\sigma_{NC}$/$\sigma_{CC}$ and 
$\eta$=$\sigma_\tau$/$\sigma_{\mu}$. 
The last ratio has to be carefully evaluated.
For istance, considering the NGS beam and $\Delta$$m^{2}$
$\simeq$1$0^{-2}$-1$0^{-3}$ e$V^{2}$, $\eta$ ranges 
between 0.25 and 0.32.
In this measurement, the systematics associated with $R_{th}$, play
of course an important role and can be in principle reduced by using a near
detector. Nevertheless, near and far beams are not identical, 
introducing a possible systematic error, requiring a detailed study.
From an experimental point of view the separation of (no $\mu$) events
from $\mu$ events is usually supposed to be straightforward. On the contrary,
this measurements become difficult at low energies, having the muons a 
shorter path length.
In NOE analysis the identification of charged and neutral current events has been 
improved by means of a neural network. The algorithm uses 24 
topological, geometrical and calorimetric event parameters as input. 
The network has been trained with $\nu_\mu$ CC and NC MonteCarlo 
events with a 
energy uniformly distributed in the range $0 \div 50 \ GeV$. 
The results of this analysis are shown in Fig.~\ref{fig:orlo}.
The $\mu$ event recognition is mainly dependent on the track length.
For neutrino energy $E_{\nu}$$\geq$5GeV a good selection/rejection efficiency  
is obtained. As expected, the efficiency decreases with the energy, 
becoming the muon track shorter.

\begin{figure}  
\vskip -3cm
\epsfig{figure=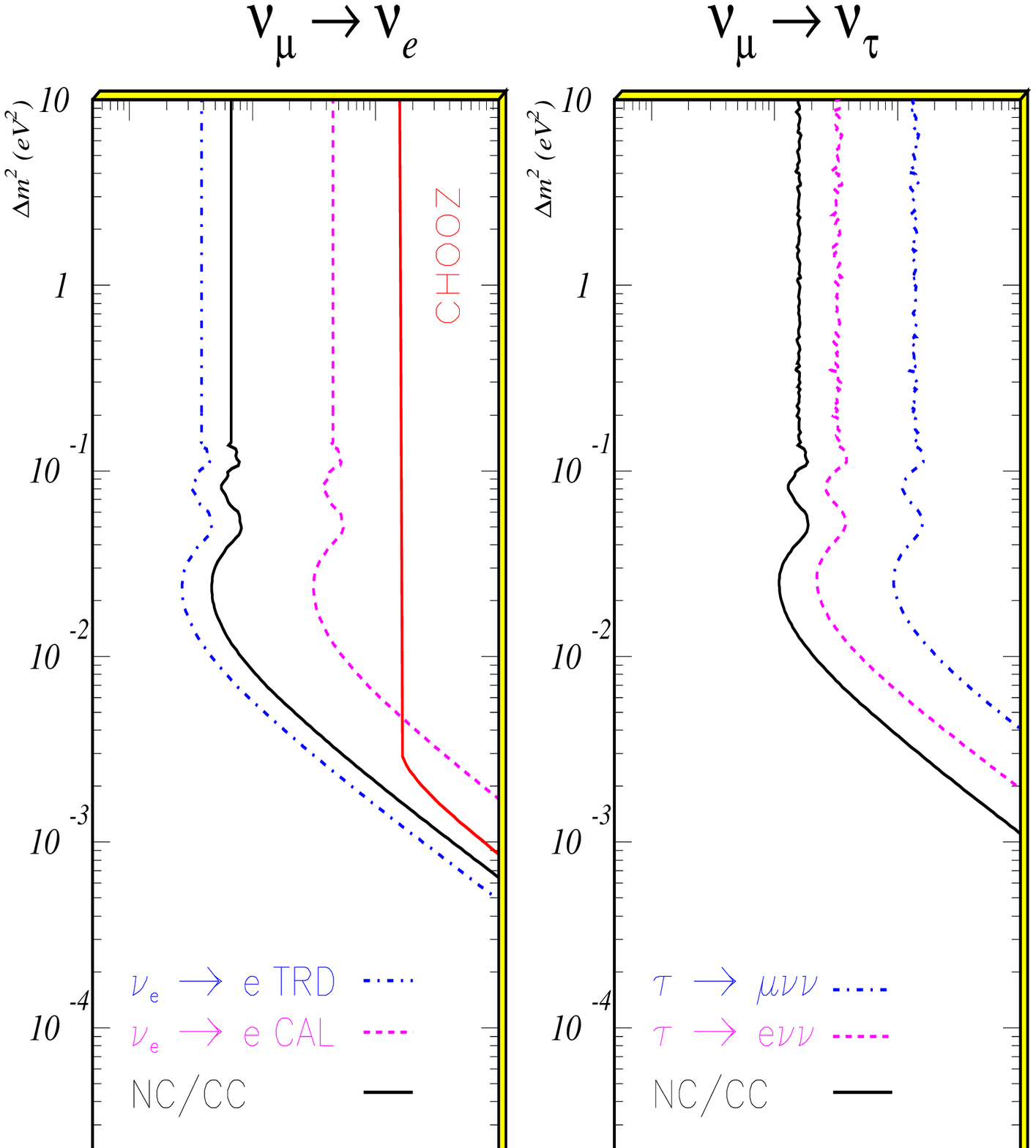,height=9cm,width=8.5cm}
\caption{\it NOE sensitivity to $\nu$ oscillations. \label{fig:finale}} 
\end{figure}
We are presently working to improve NC and CC separation at low energy.
This measurement ( Fig. 8) allows a search for $\nu_{\mu}$$\rightarrow$
$\nu_{\tau}$ oscillation down to $\Delta$$m^{2}$=1.$\cdot$1$0^{-3}$e$V^{2}$.
\section{Conclusions}
The combined use of two subdetectors (TRD and CAL) allows to search for
$\tau$ appearance signal for events generated in the $2.4 \ kton$ TRD 
target where electron identification, vertex and kinematics 
reconstruction are performed at best. Nevertheless the whole $8 \ kton$
mass can be exploited for disappearance oscillation tests.

Such measurements can be carried out at the same time by using an
appropriate neutrino beam. A full analysis has recently been performed 
to demonstrate the feasibility of both measurements. In 
Fig.~\ref{fig:finale} the sensitivity of NOE experiment to $\nu$ 
oscillations is shown.                                                 

\end{document}